\documentclass[epj]{svjour}
\usepackage{latexsym}
\usepackage[dvips,dvipdf]{graphicx}
\usepackage{epsf}
\begin{document}
\title{Cross-ladder effects in Bethe-Salpeter and Light-Front equations}
\author{J. Carbonell\inst{1} \and V.A. Karmanov\inst{2}% etc
}                     % Do not remove
\institute{Laboratoire de Physique Subatomique et Cosmologie,
 53 avenue des Martyrs,
38026 Grenoble, France \and Lebedev Physical Institute, Leninsky Prospekt 53, 119991
Moscow, Russia}
\date{}
\abstract{%Cross-ladder effects in Bethe-Salpeter (BS) and
%Light-Front (LF) equations with scalar particles are investigated.
Bethe-Salpeter (BS) equation in Minkowski space for scalar
particles is solved for a kernel given by a sum of ladder and
cross-ladder exchanges. The solution of corresponding Light-Front
(LF) equation, where we add the time-ordered stretched boxes, is
also obtained. Cross-ladder contributions are found to be very
large and attractive, whereas the influence of stretched boxes is
negligible. Both approaches -- BS and LF -- give very close
results.
 % end of PACS codes
} %end of abstract
\maketitle
%
%%%%%%%%%%%%%%%%%%%%%%%%%%%%%%%%%%%%%%%%%%%%%%%%%%%
\section{Introduction}\label{intr}

 In a preceding paper \cite{ckI} we have proposed a new
method for solving the BS equation \cite{BS} in the Minkowski
space. Our approach does not make use of the transform to the
Euclidean space (Wick rotation) and is applicable to an arbitrary
interaction kernel. This method was first applied  in \cite{ckI}
to obtain the ladder solutions of a scalar model with interaction
Lagrangian ${\cal L}_{WC}=g\phi^2\chi$. For massless exchange we
reproduced analytically the Wick-Cutkosky equation
\cite{WICK_54_CUTKOSKY_PR96_54}. For massive ladder exchange our
numerical results coincide with ones found in previous works based
on the Wick rotation.

In the present paper we solve the BS and LF equations with a
kernel given by a sum of ladder and cross-ladder graphs. This
constitutes the first calculation of cross-ladder effects in both
equations. We consider scalar constituents in the state with zero
total angular momentum. BS equation with ladder kernel was solved
in Min\-kow\-ski space in \cite{KW}.

Non ladder effects, within the same model, using
Feyn\-man-Schwinger representation, were considered in ref.
\cite{NT_PRL_96}. In this work the full set of all irreducible
cross-ladder graphs in a bound state calculation was included. In
\cite{ADT} the effect of the cross-ladder graphs  in the BS
framework  was estimated with the kernel represented through a
dispersion relation. Non ladder self-energy  effects in LF
equation have been incorporated in \cite{Ji,Adam}.

The plan of the paper is the following. In section \ref{crbk}  we sketch
the method  used in \cite{ckI} for solving the BS equation and find the
kernel  corresponding to the cross-ladder Feynman amplitude. In section
\ref{LFD} we remind the LF equation, obtain the LF cross-ladder kernel and
incorporate also two stretched box graphs. Numerical results are given in
section  \ref{num}. Section \ref{concl} contains some concluding remarks.
Technical details of the derivation are given in appendices \ref{app1},
\ref{app2} and \ref{app3}.

%%%%%%%%%%%%%%%%%%%%%%%%%%%%%%%%%%%%%%%%%%%%%%
\section{Cross-ladder kernel in BS equation}\label{crbk}

The method for solving the BS equation proposed in our previous work
\cite{ckI} is based on projecting the original equation:
\begin{eqnarray}\label{bs}
\Phi(k,p)&=&\frac{i^2}{\left[(\frac{p}{2}+k)^2-m^2+i\epsilon\right]
\left[(\frac{p}{2}-k)^2-m^2+i\epsilon\right]}
\nonumber\\
&\times&\int \frac{d^4k'}{(2\pi)^4}iK(k,k',p)\Phi(k',p)
\end{eqnarray}
on the LF plane, i.e in applying to both sides of (\ref{bs}) the integral
transform, which for the left-hand side reads
 $$
\int_{-\infty}^{\infty}\Phi(k+\beta\omega,p)d\beta
 $$
and similarly for the full right-hand side. Here $\omega$ is an
arbitrary four-vector with $\omega^2=0$. This transformation
eliminates the singularities of the BS equation. The BS amplitude
in the transformed equation is then written in terms of the
Nakanishi integral representation \cite{nakanishi}, which for zero
angular momentum reads:
\begin{eqnarray}\label{bsint}
\Phi(k,p)&=&\frac{-i}{\sqrt{4\pi}}\int_{-1}^1dz'\int_0^{\infty}d\gamma'
\\
&\times&
\frac{g(\gamma',z')}{\left[\gamma'+m^2
-\frac{1}{4}M^2-k^2-p\cdot k\; z'-i\epsilon\right]^3}
\nonumber
\end{eqnarray}
The equation satisfied by the weight function $g(\gamma,z)$
was derived in \cite{ckI} and has the form:
\begin{eqnarray}
\label{bsnew}
&&\int_0^{\infty}\frac{g(\gamma',z)d\gamma'}
{\Bigl[\gamma'+\gamma +z^2 m^2+(1-z^2)\kappa^2\Bigr]^2}=
\nonumber\\
&&
\int_0^{\infty}d\gamma'\int_{-1}^{1}dz'\;V(\gamma,z;\gamma',z')
g(\gamma',z'),
\end{eqnarray}
where $V$ is a kernel given in terms of the BS interaction kernel $iK$ by
\begin{eqnarray}\label{V}
&&V(\gamma,z;\gamma',z')=
\frac{\omega\cdot p}{\pi}\int_{-\infty}^{\infty}\frac{-iI(k+\beta
\omega,p)d\beta}
{\left[(\frac{p}{2}+k+\beta\omega)^2-m^2+i\epsilon\right]}
\nonumber\\
&&\phantom{V(\gamma,z;\gamma',z')}\times\frac{1}
{\left[(\frac{p}{2}-k-\beta\omega)^2-m^2+i\epsilon\right]},
\\
&&
\nonumber\\
&&I(k,p)=\int \frac{d^4k'}{(2\pi)^4}\frac{iK(k,k',p)}
{\left[{k'}^2+p\cdot k' z'-\gamma'-\kappa^2+i\epsilon\right]^3}.
\label{I}
\end{eqnarray}
The bound state mass $M$ enters through the parameter
 $$ \kappa^2 = m^2- \frac{1}{4}M^2. $$

Equation (\ref{bsnew}) is equivalent to the initial BS equation (\ref{bs})
and provides,  for a given kernel $K(k,k',p)$, the same bound state mass
$M$. Once $g(\gamma,z)$ is known, the BS amplitude can be restored by eq.
(\ref{bsint}). Corresponding LF wave function $\psi(k_\perp,x)$ can be
easily obtained by
\begin{equation} \label{lfwf3a}
\psi(k_\perp,x)=\frac{1}{\sqrt{4\pi}}\int_0^{\infty}\frac{x(1-x)g(\gamma',1-2x)d\gamma'}
{\Bigl[\gamma'+k_\perp^2 +m^2-x(1-x)M^2\Bigr]^2}.
\end{equation}

In \cite{ckI} for the ladder exchange
 $$
K^{(L)}(k,k',p)=\frac{-g^2}{(k-k')^2-\mu^2+i\epsilon}
 $$
we calculated the integrals (\ref{V}), (\ref{I}) for the kernel $V$ and
solved equation (\ref{bsnew}). Derivation of $V$ for the cross-ladder
kernel is quite similar but more lengthy, since the kernel itself is more
complicated.
\begin{figure}[h!]
\begin{center}
\begin{minipage}{6.cm}
\mbox{\epsfxsize=6cm\epsffile{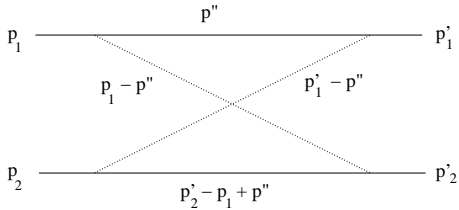}}
\end{minipage}
\end{center}
\caption{Feynman cross graph.\label{CF}}
\end{figure}
The cross-ladder BS kernel is shown in fig. \ref{CF}. On mass-shell it
depends on two variables: $s =(p_1+p_2)^2$ and $t=(p_1-p'_1)^2$. The
kernel $K$ in the BS equation (\ref{bs}) is off-mass shell. It depends
also on $p_i^2$, {\it i.e.}, on six scalar variables in general:
$s,t,p_1^2,p_2^2,{p'_1}^2,{p'_2}^2$. One can also construct six variables,
using the total momentum $p$ and relative momenta $k,k'$:
 $$ p^2,\quad
k^2,\quad {k'}^2,\quad k\cdot k',\quad k\cdot p,\quad  k'\cdot p,
 $$
where
 $$ p=p_1+p_2=p'_1+p'_2,\quad k=\frac{1}{2}(p_1-p_2), \quad
k'=\frac{1}{2}(p'_1-p'_2).
 $$
In the bound state problem the value of $p^2$ is fixed: $p^2=M^2$.

The amplitude corresponding to the diagram in fig. \ref{CF} reads:
\begin{eqnarray}\label{F1}
&&K^{(CL)}(k,k',p)=
\nonumber\\
&&\frac{-ig^4}{(2\pi)^4}\int
\frac{1}{[{p''}^2-m^2+i\epsilon]
[(p'_2-p_1+p'')^2-m^2+i\epsilon]}
\nonumber\\
&\times&
\frac{d^4p''}
{[(p_1-p'')^2-\mu^2+i\epsilon][(p'_1-p'')^2-\mu^2+i\epsilon]}.
\end{eqnarray}
We have first to calculate this expression, substitute the result in  (\ref{I}),
then in (\ref{V}) and find in this way the
cross-ladder contribution to the kernel $V(\gamma,z;\gamma',z')$ in
 equation (\ref{bsnew}).

The full kernel -- including ladder and cross-ladder graphs -- will be written
in the form:
 $$
V(\gamma,z;\gamma',z')=V^{(L)}(\gamma,z;\gamma',z')
+V^{(CL)}(\gamma,z;\gamma',z').
 $$
The ladder kernel $V^{(L)}$ was found in \cite{ckI}. The cross-ladder
contribution $V^{(CL)}$ is calculated in appendix \ref{app1}.

%%%%%%%%%%%%%%%%%%%%%%%%%%%%%%%%%%%
\section{Ladder, cross-ladder and stretched-box kernel in LF equation}\label{LFD}

We would like to compare the results obtained in the BS approach
with the equivalent ones found in Light-Front Dynamics (LFD). For
this aim we precise in what follows the LF equation and derive the
corresponding kernel. In the well-known variables
$\vec{k}_{\perp}$ and $x$ the LF equation reads (see {\it e.g.}
\cite{cdkm}):
\begin{eqnarray}\label{eq1}
&&\left(\frac{\vec{k}^2_{\perp}+m^2}{x(1-x)}-M^2\right)
\psi(\vec{k}_{\perp},x)=
\\
&&-\frac{m^2}{2\pi^3}\int\psi(\vec{k}'_{\perp},x')
V_{LF}(\vec{k}'_{\perp},x';\vec{k}_{\perp},x,M^2)
\frac{d^2k'_{\perp}dx'}{2x'(1-x')}\ .
\nonumber
\end{eqnarray}

\begin{figure}[h!]%[!ht]
\begin{center}\mbox{\epsfxsize=8.cm\epsffile{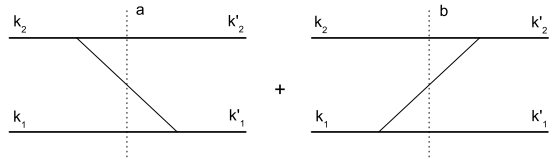}}
\caption{Ladder LFD graphs.}\label{fkern}
\end{center}
\end{figure}

The two time-ordered ladder graphs are shown in fig. \ref{fkern}. The
corresponding kernel has the form:
\begin{eqnarray}\label{lfdlad}
&&V^{(L)}_{LF}(\vec{k}'_{\perp},x';\vec{k}_{\perp},x,M^2)=
\nonumber\\
&&-\frac{4\pi\alpha\theta(x'-x)}{(x'-x)(s_a-M^2)}
-\frac{4\pi\alpha\theta(x-x')}{(x-x')(s_b-M^2)},
\end{eqnarray}
where
$$
s_a= \frac{\vec{k}^2_{\perp}+m^2}{x}
+\frac{(\vec{k'}_{\perp}-\vec{k}_{\perp})^2+\mu^2}{x'-x}+
\frac{\vec{k'}^2_{\perp}+m^2}{1-x'}
$$
and
$$
s_b= \frac{\vec{k'}^2_{\perp}+m^2}{x'}
+\frac{(\vec{k'}_{\perp}-\vec{k}_{\perp})^2+\mu^2}{x-x'}+
\frac{\vec{k}^2_{\perp}+m^2}{1-x}.
$$
\begin{figure*}[htbp]%[!ht]
\begin{center}
\begin{minipage}{8.cm}
\begin{center}
\mbox{\epsfxsize=8.cm\epsffile{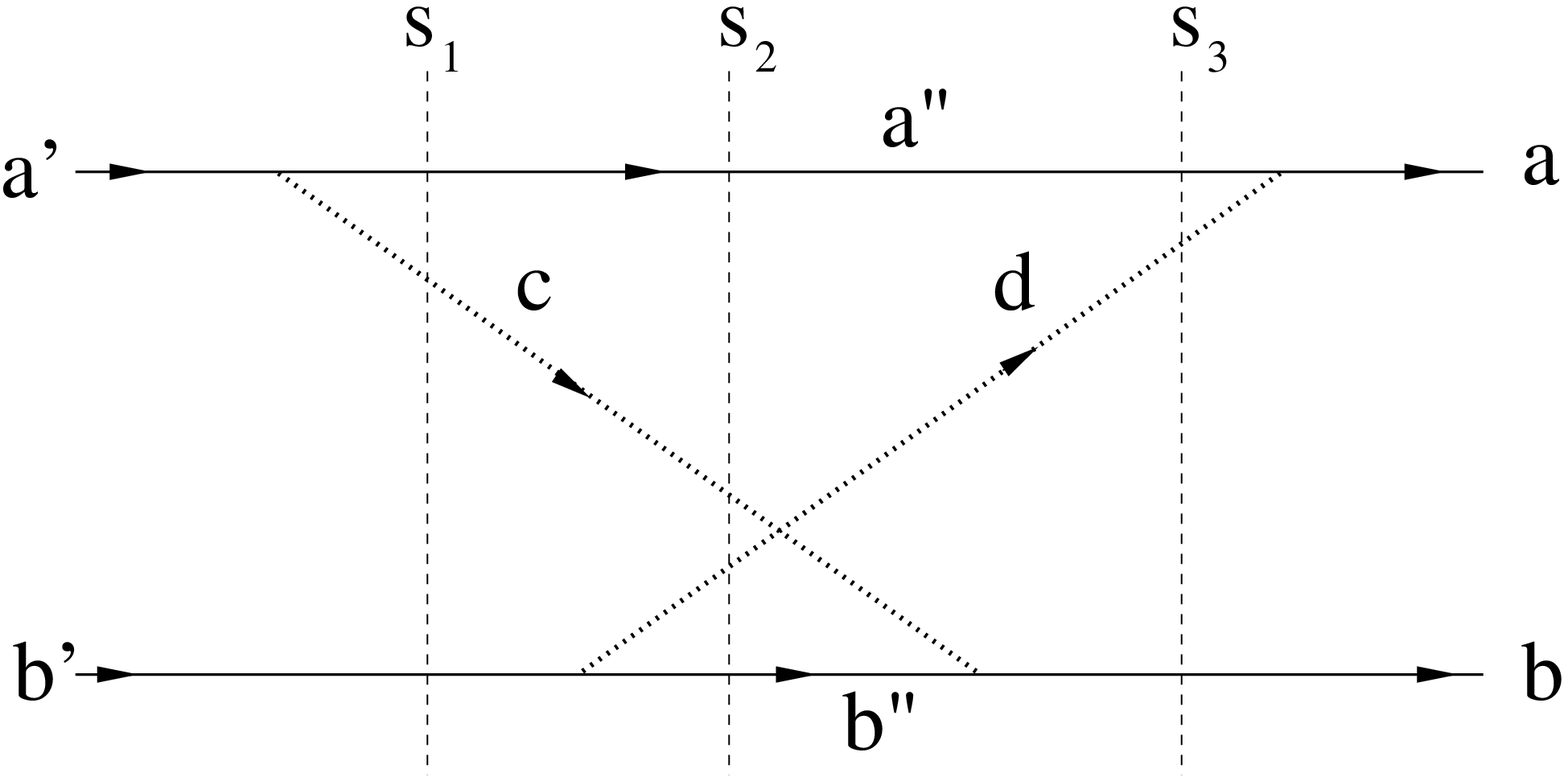}}
$V_1$
\end{center}
\end{minipage}
\hspace{1.cm}
\begin{minipage}{8.cm}
\begin{center}
\mbox{\epsfxsize=8.cm\epsffile{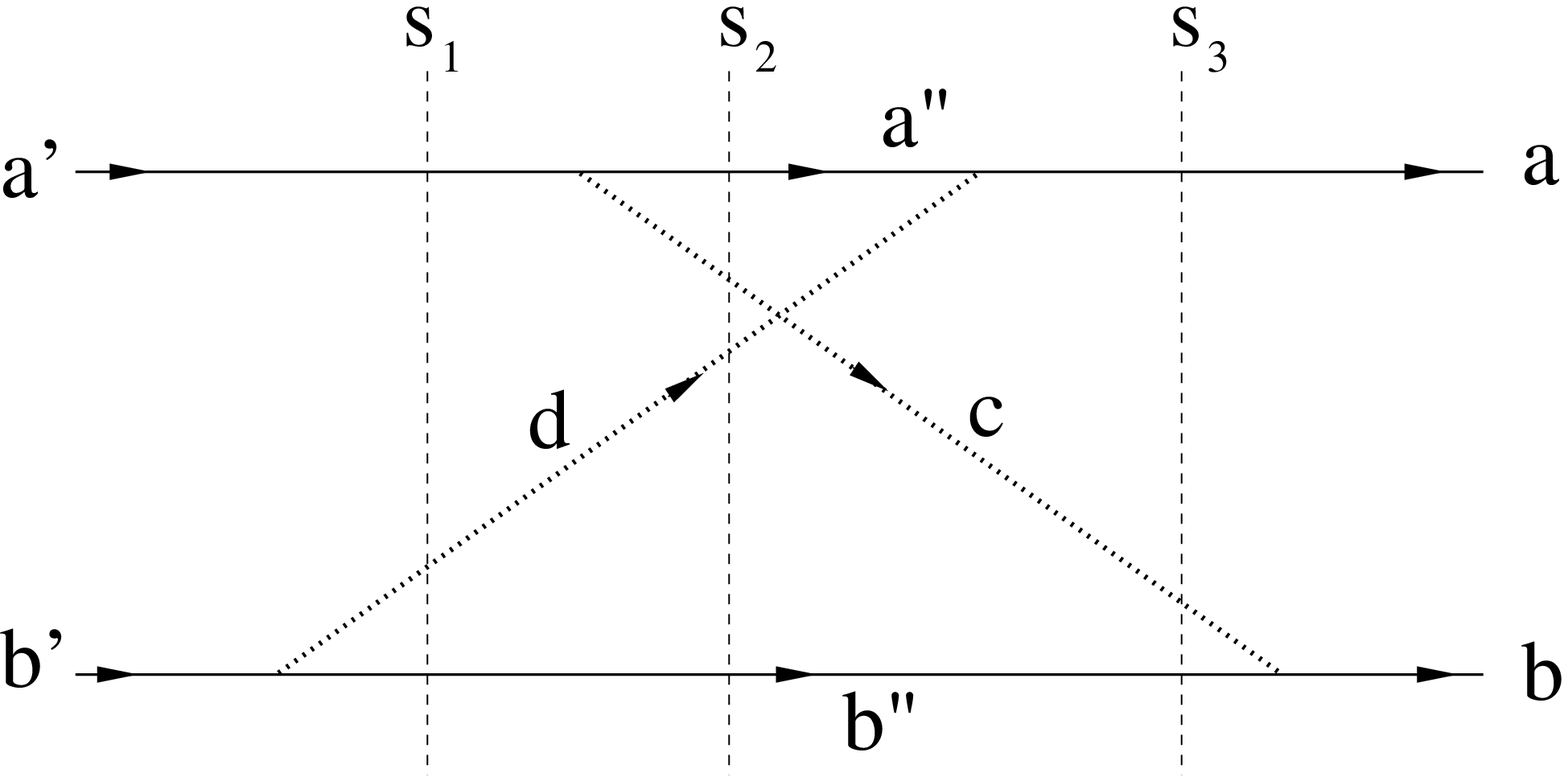}}
$V_2$
\end{center}
\end{minipage}
\end{center}
\begin{center}
\begin{minipage}{8.cm}
\begin{center}
\mbox{\epsfxsize=8.cm\epsffile{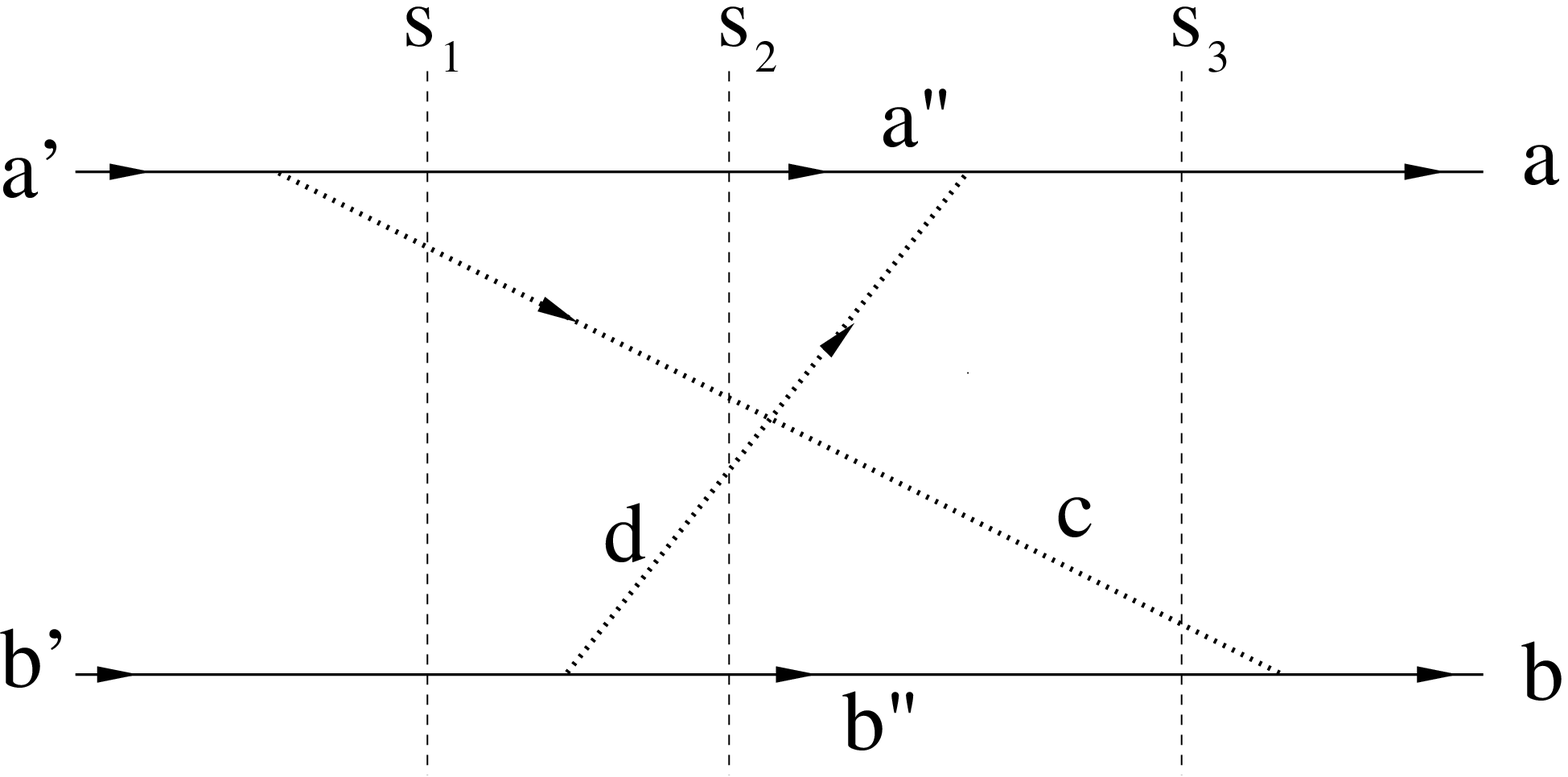}}
$V_3$
\end{center}
\end{minipage}
\hspace{1.cm}
\begin{minipage}{8.cm}
\begin{center}
\mbox{\epsfxsize=8.cm\epsffile{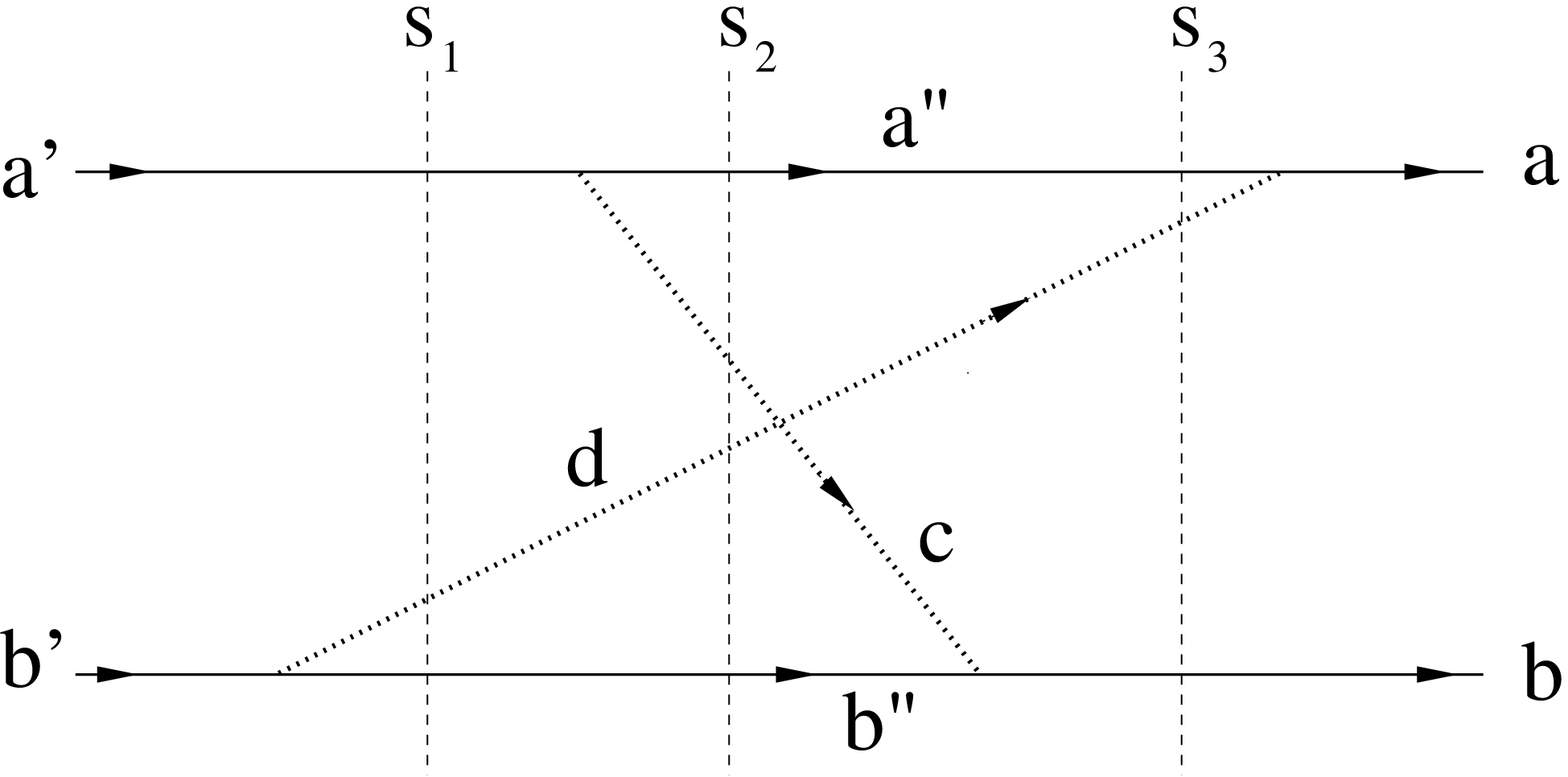}}
$V_4$
\end{center}
\end{minipage}
\end{center}
\begin{center}
\begin{minipage}{8.cm}
\begin{center}
\mbox{\epsfxsize=8.cm\epsffile{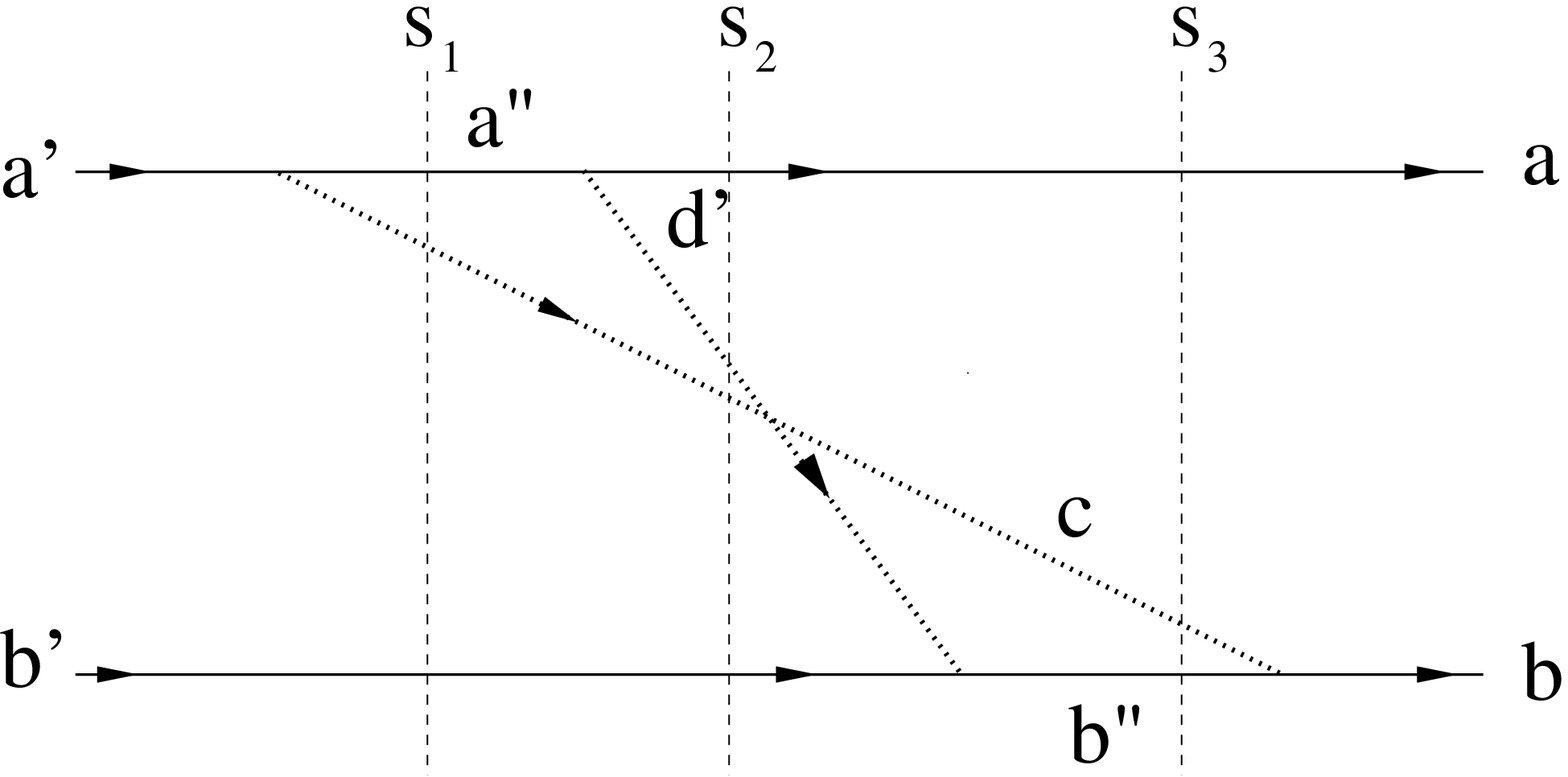}}
$V_5$
\end{center}
\end{minipage}
\hspace{1.cm}
\begin{minipage}{8.cm}
\begin{center}
\mbox{\epsfxsize=8.cm\epsffile{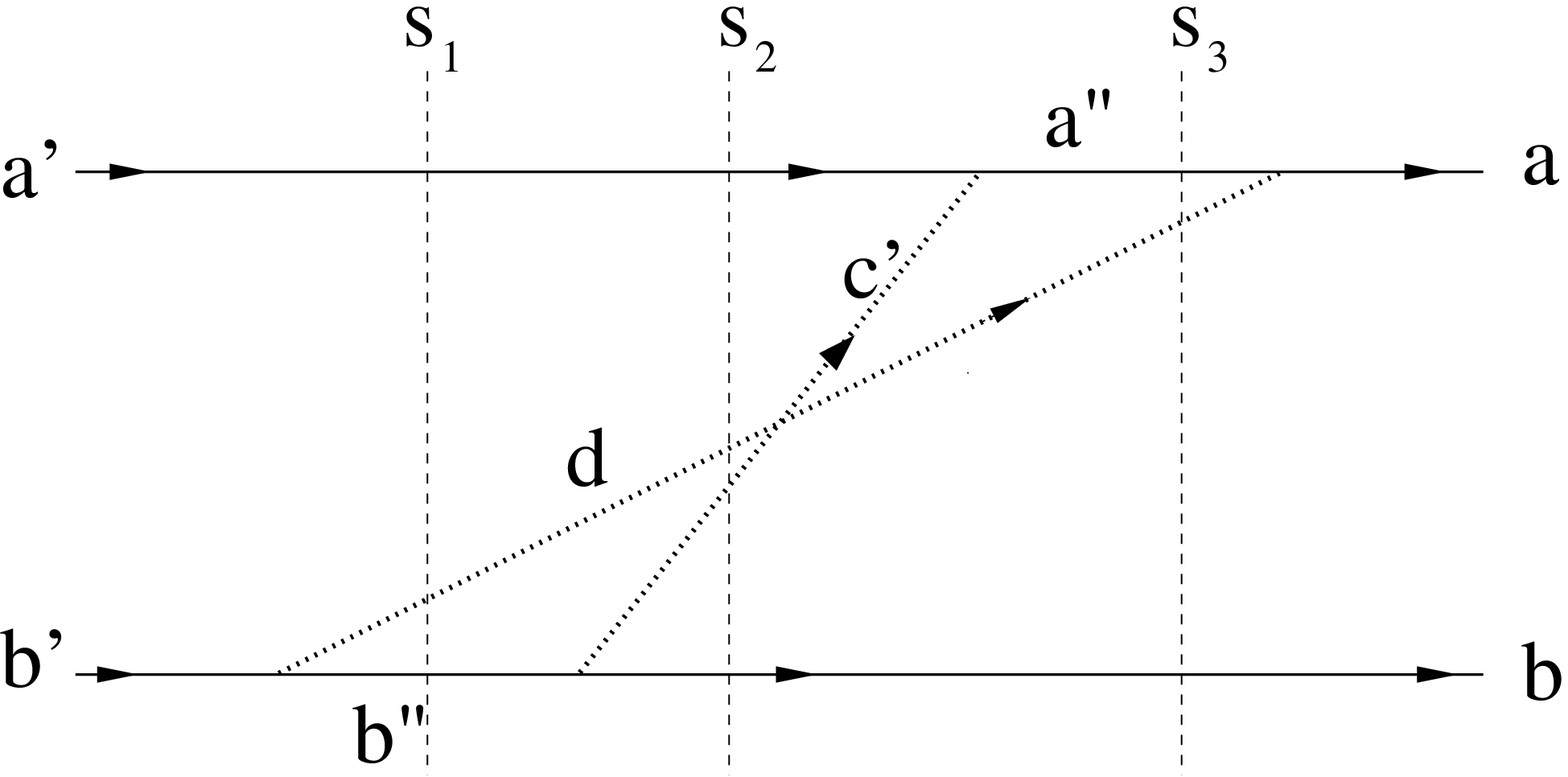}}
$V_6$
\end{center}
\end{minipage}
\end{center}
\caption{Cross LFD graphs.\label{cross}}
\end{figure*}
\begin{figure*}[htbp]%[ht!]
\begin{center}
\begin{minipage}{8.cm}
\begin{center}
\mbox{\epsfxsize=8.cm\epsffile{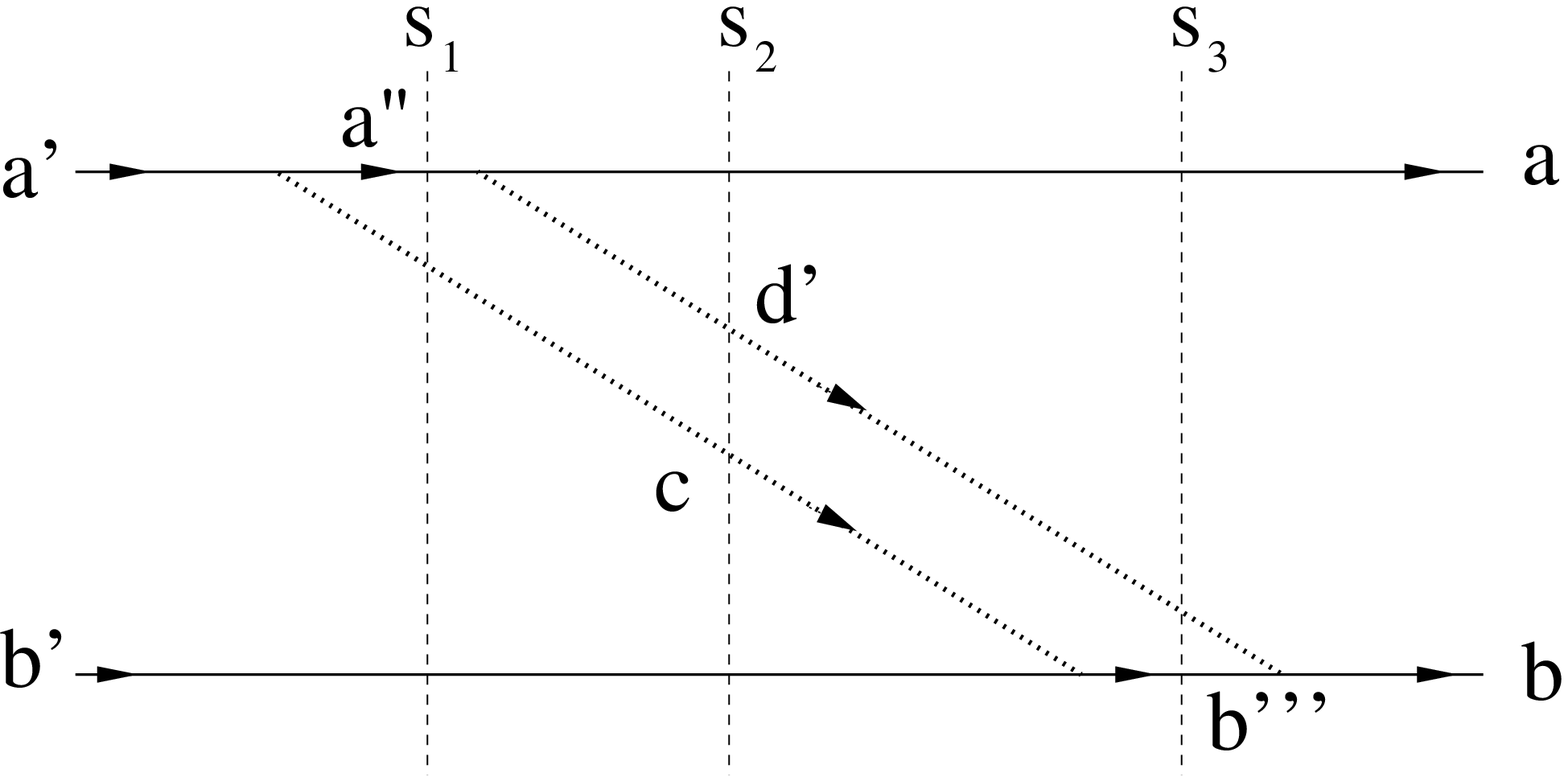}}
$V_7$
\end{center}
\end{minipage}
\hspace{1.cm}
\begin{minipage}{8.cm}
\begin{center}
\mbox{\epsfxsize=8.cm\epsffile{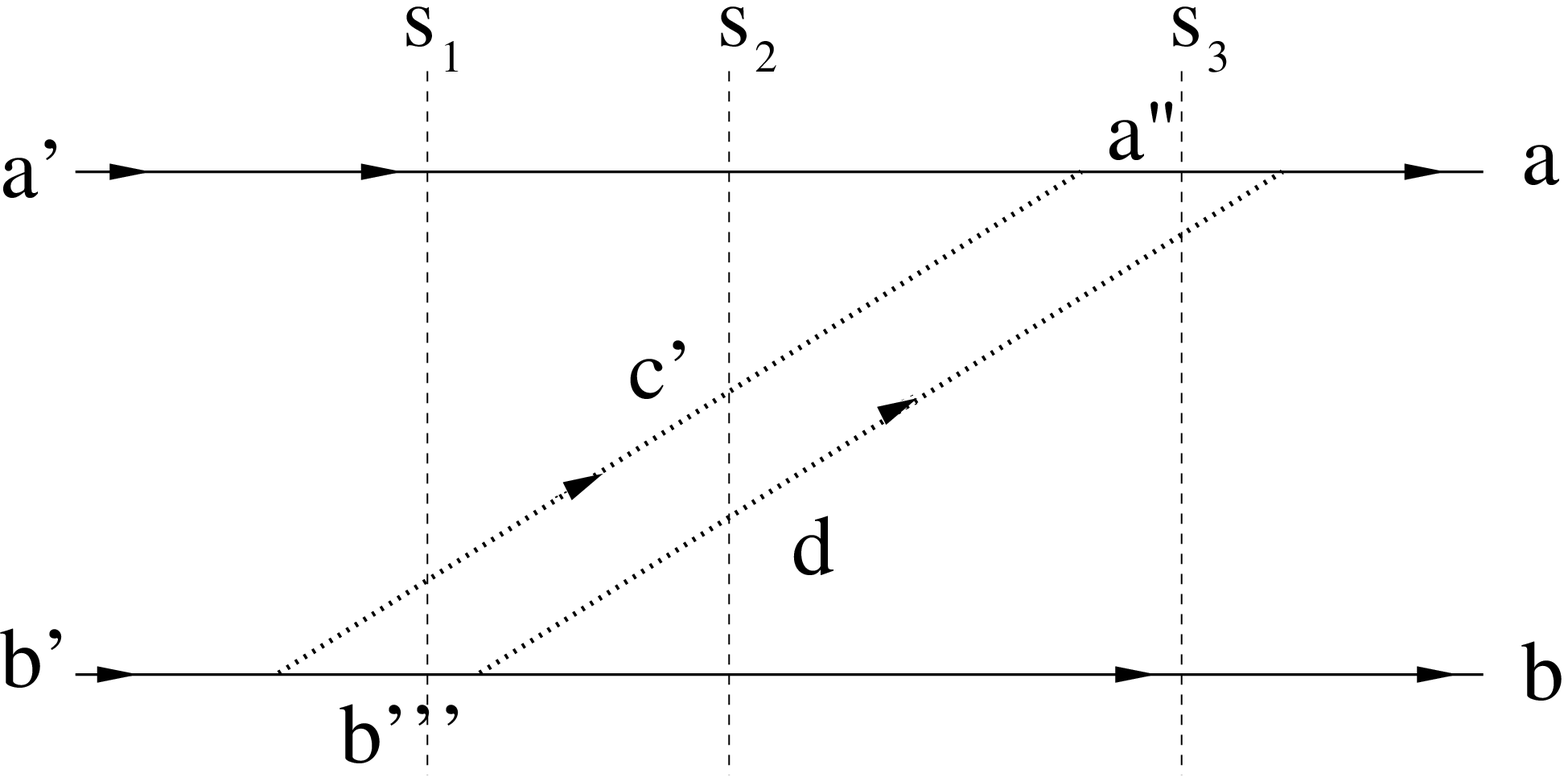}}
$V_8$
\end{center}
\end{minipage}
\end{center}
\caption{Stretched boxes.\label{box}}
\end{figure*}

The six different cross-ladder diagrams are shown in fig. \ref{cross}.
They have order $\alpha^2$. In addition, and in the same order $\alpha^2$,
there are two time-ordered irreducible graphs with two mesons in the intermediate
state (stretched boxes). They are shown in fig. \ref{box}. The full
LFD kernel -- including ladder, cross-ladder and stretched-box graphs --
will be written in the form:
 $$
V_{LF}(\vec{k}'_{\perp},x';\vec{k}_{\perp},x,M^2)=V_{LF}^{(L)}+
V_{LF}^{(CL)}+V_{LF}^{(SB)}.
 $$
The term $V_{LF}^{(L)}$ is given by (\ref{lfdlad}) and
\begin{equation}
\label{VLF}
V_{LF}^{(CL)}=\sum_{i=1}^6V_i,\quad V_{LF}^{(SB)}=\sum_{i=7,8}V_i
\end{equation}
are calculated in appendices \ref{app2} and \ref{app3} correspondingly.

%%%%%%%%%%%%%%%%%%%%%%%%%%%%%%%%%%%%%%%%%%%%%%%%%%%%%%%
\section{Numerical results}\label{num}

We have solved numerically the BS equation in the form
(\ref{bsnew}) and the LF equation (\ref{eq1}) both for the ladder
(L) and ladder+cross-ladder (L+CL) kernels. In the case of the LF
equation (\ref{eq1}) we added also the stretched box contributions
(L+CL+SB) with two intermediate mesons shown in fig. \ref{box}.
These stretched box contributions, as well as those with any
number of intermediate mesons, are generated by iterations of the
ladder BS kernel and are thus implicitly included in the BS
approach.

The numerical procedure, based in the spline expansion of the solution, is
similar to the one used in \cite{ckI}. By expanding the solution
$g(\gamma,z)$ on a spline basis we obtain a matrix equation:
\begin{equation}\label{LS}
\lambda B(M) g= A(M,\alpha)g.
\end{equation}
Like in \cite{ckI}, the matrix $B(M)$ was regularized by adding to its
diagonal a small value $\varepsilon\sim 10^{-4}\div 10^{-12}$ and
we have checked the stability of the eigenvalus relative to $\varepsilon$.
The difference with
respect to the ladder kernel is that the coupling constant $\alpha$ does
not appear linearly in the right-hand side of (\ref{LS}). For a fixed mass
$M$, the eigenvalue $\lambda$ is calculated for different values of
$\alpha$ and the value corresponding to $\lambda=1$ can be easily
extrapolated from the almost linear behaviour of $\lambda(\alpha)$.

\begin{table*}[htbp]%[!ht]
\begin{center}
\caption{Coupling constant $\alpha$ for  given values of the  binding
energy $B$ and exchanged mass $\mu=0.5$ calculated with BS and LF
equations for the ladder (L), ladder +cross-ladder (L+CL) and (in LFD) for
the ladder +cross-ladder +stretched-box (L+CL+SB) kernels.}
\label{tab1}       % Give a unique label
\begin{tabular}{|c|cc|ccc|}
\hline\noalign{\smallskip}
B & BS(L) & BS (L+CL)& LF (L) & LFD (L+CL)& LFD (L+CL+SB)\\
\noalign{\smallskip}\hline\noalign{\smallskip}
0.01 & 1.44 & 1.21 & 1.46 & 1.23 & 1.21 \\
0.05 & 2.01 & 1.62 & 2.06 & 1.65 & 1.62 \\
0.10 & 2.50 & 1.93 & 2.57 & 2.01 & 1.97 \\
0.20 & 3.25 & 2.42 & 3.37 & 2.53 & 2.47 \\
0.50 & 4.90 & 3.47 & 5.16 & 3.67 & 3.61 \\
1.00 & 6.71 & 4.56 & 7.17 & 4.97 & 4.91 \\
\noalign{\smallskip}\hline
\end{tabular}
\end{center}
\end{table*}

\begin{figure*}[htbp]%[h!]
\vspace{1cm}
\begin{center}
\begin{minipage}{12.cm}
\mbox{\epsfxsize=12.cm\epsffile{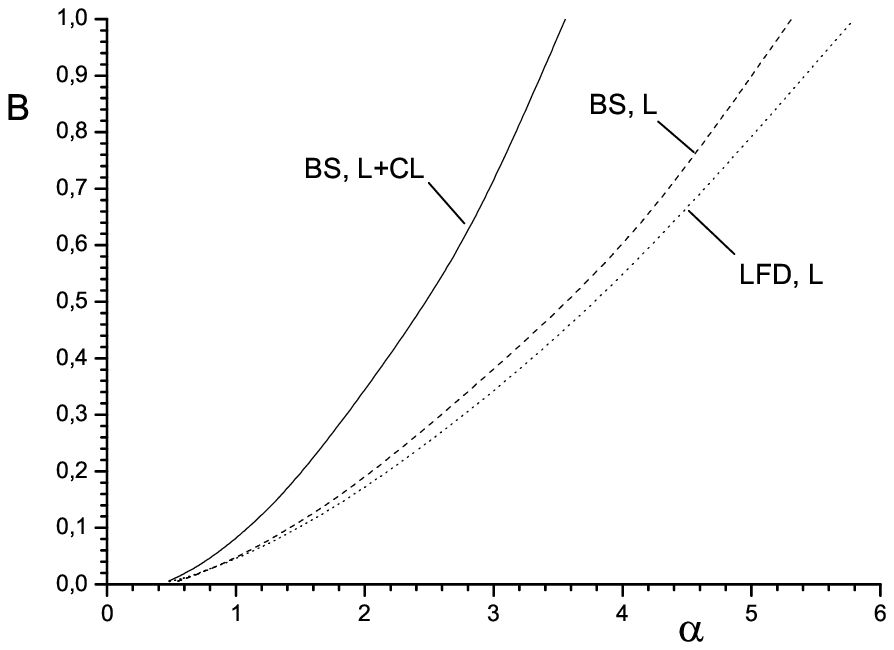}}
\end{minipage}
\end{center}
%\vspace{-1.cm}
\caption{Binding energy $B$ vs. coupling constant $\alpha$ for BS
and LF equations with the ladder (L) kernels only and with the
ladder +cross-ladder (L+CL) one for exchange mass
$\mu=0.15$.}\label{figmu015} \vspace{1.cm}
\end{figure*}

\begin{figure*}[htbp]%[h!]
\begin{center}
\begin{minipage}{12.cm}
\mbox{\epsfxsize=12.cm\epsffile{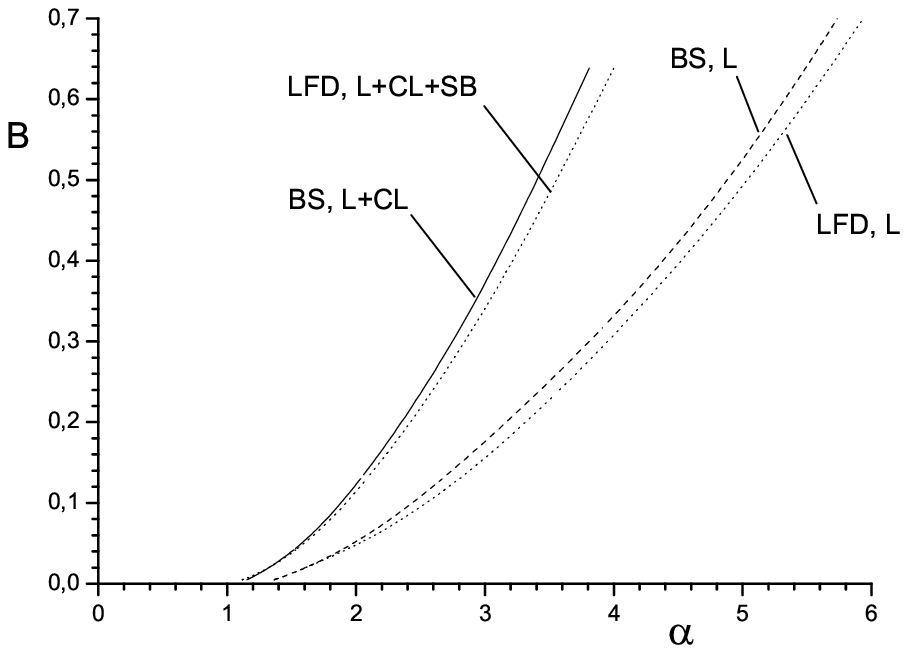}}
\end{minipage}
\end{center}
%\vspace{-1.cm}
\caption{The same as in fig. \ref{figmu015} for exchange mass
$\mu=0.5$ and, in addition, binding energy $B$ for LF equation
with the ladder +cross-ladder +stretched box (L+CL+SB)
kernel.}\label{figmu05}
\end{figure*}

The binding energy $B$ as a function of the coupling constant $\alpha$ is
shown in figures \ref{figmu015} and \ref{figmu05} for exchange masses
$\mu=0.15$ and $\mu=0.5$ respectivley and unit constituent mass ($m=1$).
Corresponding numerical values are given in table \ref{tab1}.
Obtaining this results is
quite a lengthy numerical task, entirely due to the evaluation
of the 4-dimensional integral
required in the CL kernel. This fact makes difficult -- though
straightforward -- to reach the same accuracy than for the ladder case.
The results in table \ref{tab1} have been obtained with $N\approx10$ gauss
quadrature points in each integration variable ensuring an accuracy of
1\%.

We see that for the same kernel -- ladder or (ladder +
cross-ladder) -- and exchange mass  -- $\mu=0.15$ or $\mu=0.5$ --
the binding energies obtained by BS and LFD approaches are very
close to each other. The BS equation is slightly more attractive
than LFD. At the same time, the results for ladder and (ladder
+cross-ladder) kernels considerably differ from each other. The
effect of the cross-ladder is strongly attractive. Though the
stretched box graphs are included, its  contribution to the
binding energy is smaller than 2\% and  attractive as well. This
agrees with the direct calculation of the stretched box
contribution to the kernel done in \cite{sbk}.

\begin{figure}[h!]%[htbp]%[ht!]
\centering\includegraphics[width=8cm]{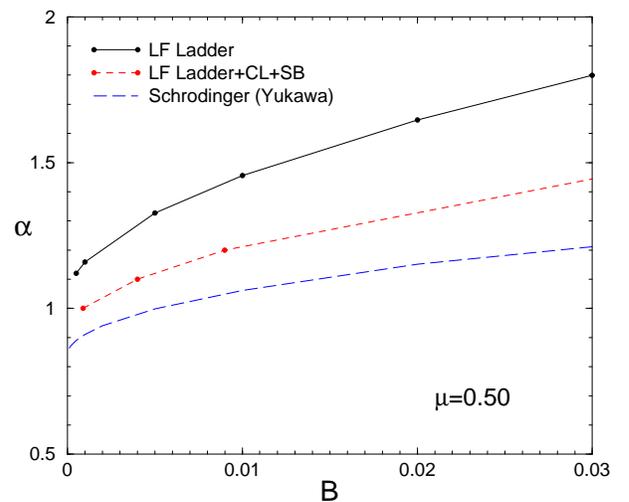}
%\vspace{-0.5cm}
\caption{Zoom of figure \ref{figmu05} in the zero binding energy region.
The ladder and (ladder +cross ladder +stretched box) results obtained with
the LF equation are compared to the non relativistic ones (Schrodinger
equation with Yukawa potential).}\label{Zoom}
\end{figure}

The zero binding limit of fig. \ref{figmu05} deserves some comments. It
was found in \cite{mariane} that for massive exchange, the relativistic
(BS and LF) ladder results do not coincide with those provided by the
Schr\"odinger equation and the corresponding non relativistic kernel
(Yukawa potential) even at very small binding energies. Their differences
increase with the exchanged mass $\mu$ and do not vanish in the limit
$B\to0$. We have displayed in fig. \ref{Zoom} a zoom of fig. \ref{figmu05}
for small values of $B$. We see from these results that the cross ladder
and stretched box diagrams reduce the differences but are not enough to
cancel it.

%%%%%%%%%%%%%%%%%%%%%%%%%%%%%%%%%%%
\section{Conclusion}\label{concl}
We have solved, for the first time, the BS equation for the kernel
given by sum of ladder and cross-ladder exchanges. The solution
was found in Minkowski space, i.e.  without making use of the Wick
rotation, by a new method developed in \cite{ckI}.

%\bigskip
In order to compare two different relativistic appro\-a\-ches, we
have also solved the corresponding LF equation.

%\bigskip
We have found that the cross-ladder contribution, relative to the
ladder one, results in a strong attractive effect. The BS and LFD
approaches give very close results for any kernel, with  BS
equation being always more attractive. These approaches differ
from each other by the stretched-box diagrams with higher numbers
of intermediate mesons. Our results indicate that the higher order
stretched box contributions are small. This agrees with direct
calculations in LFD of stretched box kernel (fig. \ref{box}) with
two-meson states \cite{sbk} and with calculations of the higher
Fock sector contributions \cite{hk04} in the Wick-Cutkosky model.
Calculation in LFD of binding energy with the stretched box
contribution (L+CL+SB) and its comparison with (L+CL) also shows
that the stretched box contribution is attractive  but small.

%\bigskip
The comparison of our results with those obtained in
\cite{NT_PRL_96}, evaluating the binding energy $B_{all}$ for the
complete set of all irreducible diagrams, shows that the effect of
the considered cross ladder graphs, though being very important,
represent only a small part of the total correction. Thus for
$\mu=0.15$ and $\alpha=0.9$ the corresponding binding energies
obtained with BS equation are $B_{L}\approx 0.035$,
$B_{L+CL}\approx 0.06$ and $B_{all}\approx 0.225$.

%\bigskip
We are aware about the only paper \cite{ADT} where the separated
effect of the cross-ladder in the BS framework has been estimated.
The method is based on writing an approximate dispersion relation
for the kernel. Our results are smaller than the ones found in
this reference by a factor 3.
%They are much more close to the
%results found in \cite{ADT} in time-ordered approach.

%%%%%%%%%%%%%%%%%%%%%%%%%%%%%%%%%%%%%%%%%%%%%%%%%%
\section*{Acknowledgements}

Numerical calculations were performed at Institut du D\'e\-ve\-loppement
et des Ressources en Informatique Scienti\-fique (IDRIS) from  CNRS. One of
the authors (V.A.K.) is sincerely grateful for the warm hospitality of the
theory group at the Laboratoire de Physique Subatomique et Cosmologie,
Universit\'e Joseph Fourier, in Grenoble, where this work was performed.
This work is supported in part  by the RFBR grant 05-02-17482-a (V.A.K.).

%%%%%%%%%%%%%%%%%%%%%%%%%%%%%%%%%%%%%%%%%%%%%%%%%%%%%%%%%%%%%
\appendix
\section{Calculating BS cross-ladder}\label{app1}
We calculate here the cross-ladder contribution $V^{(CL)}$ to the kernel
in eq. (\ref{bsnew}). We start with the cross-ladder amplitude, eq.
(\ref{F1}). Using the Feynman parametrization:
\begin{eqnarray*}
&&\frac{1}{abcd}=6\int_0^1dy_3\int_0^{1-y_3}dy_2\int_0^{1-y_2-y_3}dy_1
\\
&&\times\frac{1}{[ay_1+by_2+cy_3+d(1-y_1-y_2-y_3)]^4}
\end{eqnarray*}
and calculating then the integral (\ref{F1}) over $p''$ in a standard way
by means of the formula
\begin{equation}\label{intf}
\int \frac{d^4p''}{({p''}^2+A+i\epsilon)^n}
=\frac{i\pi^2}{(n-1)(n-2)(A+i\epsilon)^{n-2}},
\end{equation}
we find:
\begin{eqnarray}\label{K4}
&&K^{(CL)}(k,k',p)=
\\
&&16\alpha^2 m^4\int_0^1dy_3\int_0^{1-y_3}dy_2
\int_0^{1-y_2-y_3}dy_1\;\frac{1}{{A'}^2},
\nonumber
\end{eqnarray}
where $\alpha=g^2/(16\pi m^2)$ and
\begin{eqnarray}\label{D}
A'&=&{k'}^2\;(y_1+y_3)(1-y_1-y_3)
\\
&+&k^2\;(y_2+y_3)(1-y_2-y_3)
\nonumber\\
&+&k'\cdot p\;\Bigl(y_1(1-y_1-y_2)-y_3(y_1+y_2)\Bigr)
\nonumber\\
&-&k\cdot p\;
 \Bigl(y_2(1-y_1-y_2)-y_3(y_1+y_2)\Bigr)
\nonumber\\
&+&2k\cdot k'\;\Bigl((y_1+y_3)(y_2+y_3)-y_3\Bigr)
\nonumber\\
&+&\frac{1}{4}M^2\;(y_1+y_2)(1-y_1-y_2)
\nonumber\\
&-&\mu^2\;(1-y_1-y_2)-m^2\;(y_1+y_2)+i\epsilon.
\nonumber
\end{eqnarray}

Substituting the kernel (\ref{K4}) in eq. (\ref{I}), we obtain:
\begin{eqnarray}\label{Ikp}
&&I(k,p)=\int \frac{d^4k'}{(2\pi)^4}\frac{iK^{(CL})(k,k',p)}{H^3}
=
\\
&&\int_0^1dy_3\int_0^{1-y_3}dy_2\int_0^{1-y_2-y_3}dy_1
\int  \frac{d^4k'}{(2\pi)^4}\frac{16 i\alpha^2 m^4}{{A'}^2\,H^3},
\nonumber
\end{eqnarray}
where $A'$ is given by (\ref{D}) and $H$ is denominator in (\ref{I}):
$$
H={k'}^2+p\cdot k'\; z' -m^2 +\frac{1}{4}M^2 -\gamma'+i\epsilon.
$$
Using the formula
 $$ \frac{1}{{A'}^2\,H^3}=\int_0^1 \frac{12 y_4
(1-y_4)^2dy_4}{\Bigl[y_4 A'+(1-y_4) H\Bigr]^5}
 $$
and shifting the integration variable $k'$ to eliminate the terms linear
in $k'$, we calculate, again by means of eq. (\ref{intf}), the integral
(\ref{Ikp}) over $k'$. The result is represented in the form:
\begin{equation}
\label{Ikpa}
 I(k,p)=-\frac{\alpha^2 m^4}{\pi^2}
\int\frac{\eta\,y_4(1-y_4)^2dy_1dy_2dy_3dy_4}
{\Bigl[A(k,p)+i\epsilon\Bigr]^3}
\end{equation}
where
\begin{equation}\label{c2}
\eta=1-y_4[1-(1-y_1-y_3)(y_1+y_3)]
\end{equation}
and $A(k,p)$ depends on $k,p,\gamma',z',y_{1-4}$. We separated the factor
$\eta$ in numerator of (\ref{Ikpa}) so that $A(k,p)$ be a polynomial in all
the variables. We do not precise it.

Now we shift the argument $k$ of $I(k,p)$: $k\to k+\beta\omega$,
substitute $I(k+\beta\omega,p)$ in eq. (\ref{V}) for $V$ and replace
$\beta'=(\omega\cdot p)\beta$. In addition to the variables $\gamma',z'$,
which enter through eq. (\ref{I}), the kernel $V$ depends on three scalars
$k^2$, $p\cdot k$ and $\frac{\omega\cdot k}{\omega\cdot p}$: $V=V(k^2,
p\cdot k,\frac{\omega\cdot k}{\omega\cdot p};\gamma',z')$. They vary in
the intervals
\begin{equation}
\label{intervals}
-\infty < k^2 \le 0,\quad -\infty < p\cdot k < \infty,\quad
-\frac{1}{2} \le \frac{\omega\cdot k}{\omega\cdot p}\le \frac{1}{2}
\end{equation}
and satisfy the relation
\begin{equation}
\label{relat}
 p\cdot k=2\frac{\omega\cdot k} {\omega\cdot p}
\left(k^2-\kappa^2\right).
\end{equation}
Therefore, only two of them are independent. We introduce two new
variables $\gamma,z$ related to the three old ones as:
\begin{eqnarray*}
k^2&=&-\frac{(\gamma+z^2 m^2)}{1-z^2},
\\
p\cdot k&=&\frac{z[\gamma+z^2 m^2+(1-z^2)\kappa^2]}{1-z^2},
\\
\frac{\omega\cdot k}{\omega\cdot p}&=&-\frac{1}{2}z.
\end{eqnarray*}
With these definitions, the relation (\ref{relat}) turns into identity and
the inequalities (\ref{intervals}) are satisfied if $\gamma,z$ vary in the
intervals $0\le \gamma < \infty$, $-1\le z \le 1$. So, the kernel $V$ is
now parametrized as $V=V(\gamma,z;\gamma',z')$, where $\gamma$ is in the
same interval as $\gamma'$ and similarly for $z,z'$. Note that $\gamma,z$
are related to the standard LF variables $k_{\perp},x$ used in sect.
\ref{LFD} as $\gamma=k_{\perp}^2$, $z=1-2x$.

In this way we get:
\begin{eqnarray}\label{rhsc}
&&V^{(CL)}(\gamma,z;\gamma',z')=\frac{-i}{\pi(1-z^2)}
\int_{-\infty}^{\infty}I\left(k+\frac{\beta'\omega}
{\omega\cdot p},p\right)
\nonumber\\
&\times&\frac{d\beta'}
{\left(\frac{\gamma+z^2m^2}{1-z^2}+\kappa^2-\beta'-i\epsilon\right)
\left(\frac{\gamma+z^2m^2}{1-z^2}+\kappa^2+\beta'-i\epsilon\right)}.
\nonumber\\
&&
\end{eqnarray}
The polynomial $A(k,p)$ determining $I(k,p)$ has the property:
 $$
A\left(k+\frac{\beta'\omega}{\omega\cdot p},p\right)=A(k,p)
+c_{\beta'}\beta',
 $$
where $c_{\beta'}=c' y_4(1-z^2)$ and $c'$ is given in appendix \ref{app1}.
Therefore the integrand in (\ref{rhsc}) has two poles of the first order
at $\beta'=\beta_{\pm}$ with
 $$
\beta_+=+i\epsilon-\frac{\gamma+z^2m^2}{1-z^2}-\kappa^2,\quad
\beta_-=-i\epsilon+\frac{\gamma+z^2m^2}{1-z^2}+\kappa^2
$$
and one pole of the third order due to the factor
 $$\frac{1}{\left[A(k,p)+c_{\beta'}\beta'+i\epsilon\right]^3}.$$
If $c_{\beta'}>0$ ({\it i.e.}, $c'>0$), the third order pole is at
$\beta'\sim - i\epsilon$. We close the contour in the upper half-plane and
take the residue at $\beta'=\beta_+ \sim +i\epsilon$. If $c'<0$, this pole
is at $\beta'=+i\epsilon$. Then we close the contour in the lower
half-plane and take the residue at $\beta'=\beta_- \sim -i\epsilon$. The
result reads:
\begin{eqnarray*}
V^{(CL)}(\gamma,z;\gamma',z')
&=&\frac{1}{\gamma+m^2z^2+\kappa^2(1-z^2)}
\\
&\times&\left\{
\begin{array}{ll}
I(k+\frac{\omega\beta_+}{\omega\cdot p},p), & \mbox{if $c'>0$}\\
I(k+\frac{\omega\beta_-}{\omega\cdot p},p), & \mbox{if $c'<0$}
\end{array}
\right.
\end{eqnarray*}
Substituting here expression (\ref{Ikpa}) for $I$, we finally find:
\begin{eqnarray}\label{Kcl}
&&V^{(CL)}(\gamma,z;\gamma',z')=-\frac{\alpha^2
m^4}{\pi^2}\frac{(1-z^2)^3}{\gamma+m^2z^2+\kappa^2(1-z^2)}
\nonumber\\
&\times&\int_0^1 y_4(1-y_4)^2dy_4
\int_0^1dy_3\int_0^{1-y_3}dy_2\int_0^{1-y_2-y_3}
dy_1\frac{\eta}{D^3},
\nonumber\\
&&
\end{eqnarray}
$\eta$ is defined by (\ref{c2}) and $D$ reads:
\begin{eqnarray}\label{Den}
D&=&c_{\gamma}\gamma+c_{\gamma'}\gamma'
+c_{\kappa}\kappa^2+c_{m}m^2+c_{\mu}\mu^2
\nonumber\\
&-&y_4|c'|\Bigl[\gamma+m^2z^2+\kappa^2(1-z^2)\Bigr].
\end{eqnarray}
The coefficients determining $D$ are given below in  appendix \ref{app1a}.

%%%%%%%%%%%%%%%%%%%%%%%%%%%%%%%%%%%%%%%%%%%%%%%%%%%%%%%
\subsection{The coefficients determining $D$, eq. (\ref{Den})}\label{app1a}
\begin{eqnarray*}
c'&=&{y_2}^2\,\left( 1 +
     \left( -1 + y_1 + y_3
        \right) \,y_4 \right) \,
   \left( 1 + z \right)
\\
& +& y_3\,\left( {y_1}^2\,
      y_4\,\left( -1 + z \right)
 - \left( -1 + y_3 \right) \,
      \left( -1 + y_4 \right) \,
      \left( z - z' \right)
\right.
\\
&+&  \left.y_1\,\left( 1 +
        y_3\,y_4\,
         \left( -1 + z \right)  - y_4\,z +
        \left( -1 + y_4 \right) \,
         z' \right)  \right)
\\
&+&  y_2\,\left( \left( -1 +
        y_4 \right) \,\left( 1 + z \right)\right.
\\
&+&\left. {y_1}^2\,y_4\,
      \left( 1 + z \right)  +
     {y_3}^2\,y_4\,
      \left( 1 + z \right)\right.
\\
&+&\left.
     y_3\,\left( 1 + 2\,z -
        z' +
        y_4\,
         \left( -2 - 3\,z + z' \right)
        \right)\right.
\\
&+&\left. y_1\,
      \left( 1 - z' +
        y_4\,
         \left( -2 - z +
           2\,y_3\,\left( 1 + z \right)  +
           z' \right)  \right)  \right)
\\
&&
\\
c_{\gamma}&=&y_4\,\left( {y_2}^2\,
     \left( 1 + \left( -1 + y_1 +
          y_3 \right) \,y_4
       \right) \,\left( 1 + z \right)\right.
\\
&+&
    y_3\,\left( -1 + y_3 +
       \left( 1 + \left( -1 + y_1 \right)
             \,y_1 \right) \,y_4\right.
\\
       & -&\left. y_3\,y_4 +
       y_1\,y_3\,
        y_4 + y_1\,z -
       {y_1}^2\,y_4\,z -
       y_1\,y_3\,
        y_4\,z\right.
\\
&+&\left.
       \left( -1 + y_1 +
          y_3 \right) \,
        \left( -1 + y_4 \right) \,z\,
        z' \right)
\\
&+&
    y_2\,\left( -1 +
       \left( 1 + \left( -1 + y_1 \right)
             \,y_1 \right) \,y_4\right.
        - z
\\
&+& {y_3}^2\,y_4\,
        \left( 1 + z \right)%\right.
\\
&+&
       y_3\,
        \left( 2 + z +
          y_4\,
           \left( -3 - 2\,z +
             2\,y_1\,\left( 1 + z \right)
             \right)\right.
\\
&+&\left.
          \left( -1 + y_4 \right) \,z\,
           z' \right)
\\
&+&\left.\left.
       z\,\left( y_4 +
          y_1\,
           \left( 1 - z' +
             y_4\,
              \left( -2 + y_1 +
                z' \right)  \right)
          \right)  \right)  \right)
\\
&&
\\
c_{\gamma'}&=&
\left( -1 + y_4 \right) \,
  \left( -1 + \left( 1 + {y_1}^2 +
       \left( -1 + y_3 \right)\right.\right. \,
        y_3
\\
&+&\left.\left.
       y_1\,
        \left( -1 + 2\,y_3 \right)  \right)
       \,y_4 \right) \,
  \left( -1 + z^2 \right)
\\
&&
\\
c_{\kappa}&=&-\left( \left( -1 + z^2 \right) \,
    \left( -1 + {z'}^2 +
      y_4\,
       \left( 2 + \left( -1 + y_3 \right)
            \,y_3\right.\right.\right.
\\
&+&\left.\left.\left.
         y_2\,
          \left( -1 + y_2 +
            \left( -1 + y_2 +
               y_3 \right) \,z \right)\right.\right.\right.
\\
&-&
         2\,{y_1}^2\,
          \left( -1 + z' \right)  -
         y_1\,
          \left( -2 +
            y_2\,\left( 2 + z \right)\right.
\\
& +&\left.
            y_3\,\left( 2 + z \right)
            \right) \,
          \left( -1 + z' \right)
\\
&-& \left. \left. \left.
         y_3\,
          \left( 2\,y_2 +
            \left( -1 + y_2 +
               y_3 \right) \,z \right) \,
          z' - 2\,{z'}^2
         \right)\right.\right.
\\
&+&  {y_4}^2\,
       \left( -1 + y_2 + y_3 -
         {y_3}^2 +
         {y_2}^2\,
          \left( -1 + y_3 \right) \,
          \left( 1 + z \right)\right.
\\
&+&
         y_2\,
          \left( -1 + y_3 \right) \,
          \left( y_3 +
            \left( -1 + y_3 \right) \,z
            \right)%\right.\right.\right.
\\
&+& %\left.\left.\left.
         \left( -1 + y_3 \right) \,
          y_3\,z\,z' +
         y_2\,y_3\,
          \left( 2 + z \right) \,z' +
         {z'}^2
\\
&+&\left.\left.\left.
         {y_1}^2\,
          \left( -2 + y_2 +
            y_3 + y_2\,z -
            y_3\,z + 2\,z'
            \right)\right.\right.\right.
\\
&+&
         y_1\,
          \left( 2 + {y_2}^2\,
             \left( 1 + z \right)  -
            2\,z'\right.
\\
&+&
            y_3\,
             \left( -3 + y_3 -
               y_3\,z +
               \left( 2 + z \right) \,z'
               \right)%\right.\right.\right.\right.
\\
&+& \left.\left.\left.\left.
            y_2\,
             \left( -3 - 2\,z +
               2\,y_3\,
                \left( 1 + z \right)  +
               \left( 2 + z \right) \,z'
               \right)  \right)  \right)  \right)
    \right)
\\
&&
\\
c_m&=&-\left( {y_1}^3\,{y_4}^2\,
     \left( -1 + z^2 \right)  \right)
\\
&+&
  {y_2}^2\,y_4\,
   \left( 1 + \left( -1 + y_3 \right) \,
      y_4 \right) \,\left( 1 + z \right) \,
   \left( -1 + z + z^2 \right)
\\
&+&
  {y_1}^2\,y_4\,
   \left( \left( 1 + z \right) \,
      \left( -1 + z +
        y_2\,y_4\,z^2 \right)\right.
\\
 &+&\left.
      y_3\,
      \left( y_4 - y_4\,z^3
        \right)  + 2\,
      \left( -1 + y_4 \right) \,
      \left( -1 + z^2 \right) \,z' \right)
\\
&+& y_2\,y_4\,
   \left( z^2\,\left( -1 + y_4 +
        y_3\,
         \left( 2 + \left( -3 + y_3 \right)
              \,y_4 \right)\right.\right.
\\
&-&\left.\left. z +
        \left( y_3 +
           {\left( -1 + y_3 \right) }^2\,
            y_4 \right) \,z \right)\right.
\\
&+&\left.
     y_3\,\left( -1 + y_4
        \right) \,\left( -2 +
        z^2\,\left( 2 + z \right)  \right) \,
      z' \right)
\\
 &+&
  \left( -1 + y_4 \right) \,
   \left( -\left( \left( -1 + y_3 \right)
          \,y_3\,y_4\,z^2
        \right)\right.
\\
&+& \left.\left( -1 + y_3 \right)
        \,y_3\,y_4\,z^3\,
      z' +
     \left( -1 + y_4 \right) \,
      \left( -1 + z^2 \right) \,{z'}^2
     \right)
\\
 &+& y_1\,y_4\,
   \left( -\left( y_3\,
        \left( y_4\,
           \left( 1 +
             y_3\,\left( -1 + z \right)
             \right)  - z \right) \,z^2 \right)\right.
\\
&+&
     \left.{y_2}^2\,y_4\,
      \left( 1 + z \right) \,
      \left( -1 + z + z^2 \right)\right.
\\
&+&%\left.
     \left( -1 + y_4 \right) \,
      \left( 2 - 2\,z^2 +
        y_3\,
         \left( -2 + z^2\,\left( 2 + z \right)
           \right)  \right) \,z'
\\
&+&\left.
     y_2\,\left( -\left( \left( -2 +
             z^2\,\left( 2 + z \right)  \right) \,
           \left( -1 + z' \right)  \right)\right.\right.
\\
            &+& %\left.\left.
         y_4\,
         \left( 2 - 2\,z' +
           z^2\,\left( -3 - 2\,z +
              2\,y_3\,
               \left( 1 + z \right)\right.\right.
\\
&+&\left.\left.\left.\left.
              \left( 2 + z \right) \,z'
              \right)  \right)  \right)  \right)
\\
&&
\\
c_{\mu}&=&\left( -1 + y_1 + y_2 \right) \,
  y_4\,\left( -1 +
    \left( 1 + {y_1}^2 +
       \left( -1 + y_3 \right) \,
        y_3\right.\right.
\\
&+&\left.\left.
       y_1\,
        \left( -1 + 2\,y_3 \right)  \right)
       \,y_4 \right) \,
  \left( -1 + z^2 \right)
\end{eqnarray*}

%%%%%%%%%%%%%%%%%%%%%%%%%%%%%%%%%%%%%%%%%%%%%%%%%%%
\section{Calculating LFD cross-ladders}\label{app2}
At first we will transform the 3-dim. LF equation
(\ref{eq1}) in a 2-dim. form. The kernel
$V_{LF}(\vec{k}'_{\perp},x';\vec{k}_{\perp},x,M^2)$
depends on the scalar product
 $$
\vec{k}_{\perp}\cdot\vec{k'}_{\perp}=k_{\perp}k'_{\perp}\cos\phi'.
 $$
Since $\psi(\vec{k'}_{\perp},x)$ depends on $|\vec{k'}_{\perp}|\equiv
{k'}_{\perp}$ and does not depend on the angle $\phi'$, one can integrate
the kernel over $\phi'$ and the equation (\ref{eq1}) turns into:
\begin{eqnarray}\label{eq1a}
&&\left(\frac{k^2_{\perp}+m^2}{x(1-x)}-M^2\right)
\psi(k_{\perp},x)=
\\
&&-\frac{m^2}{2\pi^3}\int\psi(k'_{\perp},x')
\tilde{V}_{LF}(k'_{\perp},x',k_{\perp},x,M^2)
\frac{k'_{\perp}dk'_{\perp}dx'}{2x'(1-x')},
\nonumber
\end{eqnarray}
where
\begin{equation}
\label{Vt}
\tilde{V}_{LF}(k'_{\perp},x',k_{\perp},x,M^2)=
\int_0^{2\pi}V_{LF}(\vec{k'}_{\perp},x',\vec{k}_{\perp},x,M^2)d\phi'.
\end{equation}

To calculate the kernel in the LF equation (\ref{eq1}), one can use the
Weinberg rules \cite{sw}, which are equivalent to the graph techniques in
LFD \cite{cdkm}. To calculate the amplitude $-{\cal M}$,  one should put
in correspondence: to every vertex -- the factor $g$, to every
intermediate state -- the factor,
 $$ \frac{2}{s_0-s_{int}+i0},\quad \mbox{where}\quad
s_{int}=\sum_i \frac{\vec{k}^2_{i\perp}+m_i^2}{x_i}
 $$
and $s_0$ is the initial (=final) state energy. In our case (bound state):
$s_0^2=M^2$. To every internal line one should put in correspondence the
factor $ \frac{\theta(x_i)}{2x_i}.$ One should take into account the
conservation laws for $\vec{k}_{i\perp}$ and $x_i$ in any vertex and
integrate over all independent variables with the measure
$\frac{d^2k_{i\perp}dx_i}{(2\pi)^3}$.

Applied to the ladder graphs fig. \ref{fkern}, these rules result in
eq. (\ref{lfdlad}) for the ladder contribution. The integral (\ref{Vt})
is calculated analytically.

There are six cross-ladder time-ordered diagrams shown in fig.
\ref{cross}. Sum of them determines, by eq. (\ref{VLF}), the kernel
$V_{LF}^{(CL)}$. Consider four diagrams for  $V_{1-4}$ in fig.
\ref{cross}. They contain eight lines ($a',a'',a,b',b'',b,c,d$) and three
intermediate states. The momenta corresponding to these lines are the
following:
\begin{eqnarray*}%\label{mom}
&&(a')\quad \vec{k'}_{\perp},\;x' \nonumber\\
&&(a'')\quad\vec{k''}_{\perp},\;x'' \nonumber\\
&&(a)\quad\vec{k}_{\perp},\;x \nonumber\\ &&(b')\quad
-\vec{k'}_{\perp},\;1-x' \nonumber\\
&&(b'')\quad\vec{k''}_{\perp}-\vec{k'}_{\perp}-\vec{k}_{\perp},\;1+x''-x'-x
\nonumber\\ &&(b)\quad -\vec{k}_{\perp},\;1-x \nonumber\\
&&(c)\quad\vec{k'}_{\perp}-\vec{k''}_{\perp},\;x'-x'' \nonumber\\
&&(d)\quad\vec{k}_{\perp}-\vec{k''}_{\perp},\;x-x''
\end{eqnarray*}
Introduce for every line the energy:
\begin{eqnarray*}
E_{a'}&=&\frac{\vec{k'}^2_{\perp}+m^2}{x'} \nonumber\\
E_{a''}&=&\frac{\vec{k''}^2_{\perp}+m^2}{x''}\nonumber\\
E_{a}&=&\frac{\vec{k}^2_{\perp}+m^2}{x}\nonumber\\
E_{b'}&=&\frac{\vec{k'}^2_{\perp}+m^2}{1-x'} \nonumber\\
E_{b''}&=&\frac{(\vec{k''}_{\perp}-\vec{k'}_{\perp}-\vec{k}_{\perp})^2+m^2}{1+x''-x'-x}\nonumber\\
E_{b}&=&\frac{\vec{k}^2_{\perp}+m^2}{1-x}\nonumber\\
E_{c}&=&\frac{(\vec{k'}_{\perp}-\vec{k''}_{\perp})^2+\mu^2}{x'-x''};\nonumber\\
E_{d}&=&\frac{(\vec{k}_{\perp}
-\vec{k''}_{\perp})^2+\mu^2}{x-x''}.
\end{eqnarray*}
Energies in three intermediate states for $V_1$ are written as:
\begin{eqnarray*}
s_1&=&E_{a''}+E_{c}+E_{b'},\\
s_2&=&E_{a''}+E_c+E_d+E_{b''},\\
s_3&=&E_{a''}+E_d+E_b.
\end{eqnarray*}

The kernel $V_{LF}$ is related to the amplitude ${\cal M}$ as:
$V_{LF}=-{\cal M}/(4m^2)$. For the diagram fig. \ref{cross},$V_1$ it has
the form:
\begin{eqnarray}\label{V1}
V_1&=&-\int \frac{g^4\theta(x'')\theta(x-x'')\theta(x'-x'')
\theta(1+x''-x-x')}
{4m^2(s_1-M^2)(s_2-M^2)(s_3-M^2)}
\nonumber\\
&\times&\frac{1}{
x'' (x-x'')(x'-x'')(1+x''-x-x')}\frac{d^2k''_{\perp}dx''}{2(2\pi)^3}.
\nonumber\\
&&
\end{eqnarray}
Since:
\begin{eqnarray*}
\vec{k}_{\perp}\cdot\vec{k'}_{\perp}&=&k_{\perp}k'_{\perp}\cos\phi',\\
\vec{k}_{\perp}\cdot\vec{k''}_{\perp}&=&k_{\perp}k''_{\perp}\cos\phi'',\\
\vec{k'}_{\perp}\cdot\vec{k''}_{\perp}&=&k'_{\perp}k''_{\perp}\cos(\phi'-\phi''),
\end{eqnarray*}
the integrand depends on two azimuthal angles $\phi',\phi''$. The kernel
(\ref{V1}) is determined by the 3-dim. integral over
$d^2k''_{\perp}dx''=k''_{\perp}dk''_{\perp}d\phi''dx''$. Therefore the
contribution of fig. \ref{cross},$V_1$ to the kernel $\tilde{V}_{LF}$, eq.
(\ref{Vt}), is determined by a 4-dim. integral:
\begin{eqnarray}\label{M3}
&&\tilde{V}_1(k'_{\perp},x',k_{\perp},x,M^2)=
\int_0^{2\pi}V_1(\vec{k'}_{\perp},x',\vec{k}_{\perp},x,M^2)d\phi'
\nonumber\\
&=&\int_{max(0,x+x'-1)}^{min(x,x')}dx''\int
\frac{-4m^2\alpha^2k''_{\perp}dk''_{\perp}d\phi'' d\phi'}
{\pi(s_1-M^2)(s_2-M^2)(s_3-M^2)}
\nonumber\\
&\times&\frac{1}{x''
(x-x'')(x'-x'')(1+x''-x-x')}
\end{eqnarray}
with $\alpha=g^2/(16\pi m^2)$. We have removed the theta-func\-ti\-ons and
incorporated the integration limits in the variable $x''$ explicitly. One
can calculate the 4-dim. integral (\ref{M3}) for the kernel numerically.

Consider now another three kernels $V_2,V_3,V_4$ shown in figs.
\ref{cross}. They still have the form (\ref{M3}), but the corresponding
energies $s_1,s_2,s_3$ are different.

We denote the kernel corresponding to fig. \ref{cross}, $V_2$, after
integration over $\phi'$, as $\tilde{V}_2(k'_{\perp},x',k_{\perp},x,M^2)$.
It has exactly the same form as eq. (\ref{M3}), but with energies given
by:
\begin{eqnarray*}
s_1&=&E_{a'}+E_{d}+E_{b''},\\
s_2&=&E_{a''}+E_c+E_d+E_{b''},\\
s_3&=&E_{a}+E_c+E_{b''}.
\end{eqnarray*}

The kernel $\tilde{V}_3(k'_{\perp},x',k_{\perp},x,M^2)$ corresponding to
the graph fig. \ref{cross}, $V_3$ has exactly the same form as eq.
(\ref{M3}), but with energies given by:
\begin{eqnarray*}
s_1&=&E_{a''}+E_{c}+E_{b'},\\
s_2&=&E_{a''}+E_c+E_d+E_{b''},\\
s_3&=&E_{a}+E_c+E_{b''}.
\end{eqnarray*}

The kernel $\tilde{V}_4(k'_{\perp},x',k_{\perp},x,M^2)$ corresponding to
the graph fig. \ref{cross}, $V_4$ also has exactly the same form as eq.
(\ref{M3}), but with energies given by:
\begin{eqnarray*}
s_1&=&E_{a'}+E_{d}+E_{b''},\\
s_2&=&E_{a''}+E_c+E_d+E_{b''},\\
s_3&=&E_{a''}+E_d+E_{b}.
\end{eqnarray*}

Another  two cross-ladder graphs are shown in figs. \ref{cross}, $V_5$ and
$V_6$. Here we have new lines $d',c'$. Corresponding momenta and energies
are the following:
\begin{eqnarray*}
&&(d')\quad \vec{k''}_{\perp}-\vec{k}_{\perp},\;x''-x
\nonumber\\
&&(c')\quad \vec{k''}_{\perp}-\vec{k'}_{\perp},\;x''-x'
\end{eqnarray*}

\begin{eqnarray*}
E_{d'}&=&\frac{(\vec{k''}_{\perp}-\vec{k}_{\perp})^2+\mu^2}{x''-x}=-E_d\ ,
\nonumber\\
E_{c'}&=&\frac{(\vec{k''}_{\perp}-\vec{k'}_{\perp})^2+\mu^2}{x''-x'}=-E_c\ .
\end{eqnarray*}

If $x'>x$, only $V_5$ contributes. It has the form:
\begin{eqnarray}\label{Vcr5}
\tilde{V}_5&=&\int_x^{x'}dx''\int
\frac{-4m^2\alpha^2\; d^2k''_{\perp} d\phi'}
{\pi(s_1-M^2)(s_2-M^2)(s_3-M^2)}
\nonumber\\
&\times&\frac{\theta(x'-x)}{ x'' (x''-x)(x'-x'')(1+x''-x-x')}
\end{eqnarray}
with $s_1,s_2,s_3$ given by:
\begin{eqnarray*}
s_1&=&E_{a''}+E_{c}+E_{b'},\\
s_2&=&E_{a}+E_c+E_{d'}+E_{b'},\\
s_3&=&E_{a}+E_c+E_{b''}.
\end{eqnarray*}

If $x>x'$, only  $V_6$ contributes. It has the form:
\begin{eqnarray}\label{Vcr6}
\tilde{V}_6&=&\int_{x'}^{x}dx''\int
\frac{-4m^2\alpha^2 d^2k''_{\perp} d\phi'}
{\pi(s_1-M^2)(s_2-M^2)(s_3-M^2)}
\nonumber\\
&\times&\frac{\theta(x-x')}{x'' (x-x'')(x''-x')(1+x''-x-x')}
\end{eqnarray}
with $s_1,s_2,s_3$ given by:
\begin{eqnarray*}
s_1&=&E_{a'}+E_{d}+E_{b''},\\
s_2&=&E_{a'}+E_{c'}+E_{d}+E_{b},\\
s_3&=&E_{a''}+E_{d}+E_{b}.
\end{eqnarray*}

%%%%%%%%%%%%%%%%%%%%%%%%%%%%%%%%
\subsection{Test by the Feynman cross graph}\label{test}
On the energy shell the sum of six contributions $V^{(CL)}= \sum_{i=1}^6
V_i$, fig. \ref{cross}, must coincide with the value of the  Feynman cross
graph, fig. \ref{CF}, taken on the mass shell, namely, with
 $$
V_{Feyn}^{(CL)}=-\frac{1}{4m^2}K^{(CL)}(k,k',p).
 $$
$K^{(CL)}(k,k',p)$ is given by (\ref{K4}) where in the denominator $A'$,
eq. (\ref{D}), one should put
\begin{eqnarray*}
&&M^2=s,\quad k^2={k'}^2=m^2-\frac{1}{4}s,\\
&&k\cdot p=k'\cdot p=0,\quad
k\cdot k'= m^2-\frac{1}{4}s-\frac{1}{2}t
\end{eqnarray*}
and $s,t$ are in the physical domain. Expression (\ref{D}) is considerably
simplified and we find:
\begin{eqnarray}\label{F4}
V_{Feyn}^{(CL)}&=&-4\alpha^2m^2
\int_0^1 dy_3\int_0^{1-y_3}dy_2\int_0^{1-y_2-y_3}dy_1
\nonumber\\
&\times&
\Bigl[(1-y_1-y_2)(\mu^2-y_3t)+(y_1-y_2)^2m^2
\nonumber\\
&& +y_3^2t+y_1y_2(s+t)\Bigr]^{-2}.
\end{eqnarray}
The 3-dim. integral (\ref{F4}) is calculated numerically. One can chose a
particular kinematics, for example, in c.m.-frame, where $s$ and $t$ for
elastic scattering are determined by incident (=final) momentum and the
scattering angle.

On the other hand, we calculate the sum of six LF cross graphs
 $$V_{LF}^{(CL)}(\vec{k'}_{\perp},x',\vec{k}_{\perp},x,M^2=s)
=\sum_{i=1}^6 V_i
 $$
(not integrated  over $\phi'$). Here
 $$ x=\frac{\omega\cdot
p_1}{\omega\cdot (p_1+p_2)},\quad x'=\frac{\omega\cdot p'_1}{\omega\cdot
(p_1+p_2)}
 $$
and $\vec{k}_{\perp},\vec{k'}_{\perp}$ are projections of $\vec{p}_1$,
$\vec{p'}_1$ on the plane orthogonal to $\vec{\omega}$. We remind that
$\omega=(\omega_0,\vec{\omega})$ is an arbitrary four-vector with
$\omega^2=0$. Though, for a given kinematics, the values of
$\vec{k}_{\perp},x,\vec{k'}_{\perp},x'$ depend on  orientation of
$\vec{\omega}$, the on-shell amplitude
$V_{LF}^{(CL)}(\vec{k'}_{\perp},x',\vec{k}_{\perp},x,M^2=s)$ does not depend on
it. We can chose $\omega=(1,0,0,-1)$, find the arguments of
$V_{LF}^{(CL)}(\vec{k'}_{\perp},x',\vec{k}_{\perp},x,M^2=s)$ in kinematics for
the c.m. elastic scattering and calculate $V_{LF}^{(CL)}$.

For different incident momenta and scattering angles we have checked
numerically that  $V_{Feyn}^{(CL)}$, eq. (\ref{F4}), coincides with the sum of
six cross graphs $V_{LF}^{(CL)}= \sum_{i=1}^6 V_i$. This confirms validity of
both calculations.

%%%%%%%%%%%%%%%%%%%%%%%%%%%%%%%%%%%%%%%%%%%%%%%%%%%%
\section{Calculating LFD stretched boxes}\label{app3}
The stretched-box graphs determine the kernel $V_{LF}^{(SB)}$, eq.
(\ref{VLF}). They are shown in fig. \ref{box} and their contributions are
denoted as $V_7$ and $V_8$. We introduce the line $b'''$. All the momenta
except for the line $b'''$ are the same as for the cross graphs. The
momentum and energy carried by the lines $b'''$ are the following:
\begin{eqnarray*}
&& (b''')\quad -\vec{k''}_{\perp},\;1-x''
\\
&&E_{b'''}=\frac{\vec{k''}^2_{\perp}+m^2}{1-x''}
\end{eqnarray*}

The following product of the theta-functions enters the kernel $V_7$
 $$
\theta(x'')\theta(x''-x)\theta(x'-x'')\theta(1-x'').
 $$
It restricts the integration domain as: $$0\leq x\leq x'' \leq x'\leq 1.$$
Therefore the kernel $\tilde{V}_7$ is nonzero at $x'>x$ only. It has the
form:
\begin{eqnarray}\label{V5}
\tilde{V}_7&=&\int_x^{x'}dx''\int
\frac{-4m^2\alpha^2\;d^2k''_{\perp} d\phi'}
{\pi(s_1-M^2)(s_2-M^2)(s_3-M^2)}
\nonumber\\
&\times&
\frac{\theta(x'-x)}{ x'' (x''-x)(x'-x'')(1-x'')}
\end{eqnarray}
The corresponding energies $s$ read:
\begin{eqnarray*}
s_1&=&E_{a''}+E_{c}+E_{b'},\\
s_2&=&E_{a}+E_{d'}+E_c+E_{b'},\\
s_3&=&E_{a}+E_{d'}+E_{b'''}
\end{eqnarray*}
and the energies $E$ are given above in appendix \ref{app2} and in this
appendix.

The kernel $\tilde{V}_8$ is non-zero at $x'<x$ only. It has the form:
\begin{eqnarray}\label{V6}
\tilde{V}_8&=&\int_{x'}^{x}dx''\int
\frac{-4m^2\alpha^2\;d^2k''_{\perp} d\phi'}
{\pi(s_1-M^2)(s_2-M^2)(s_3-M^2)}
\nonumber\\
&\times&
\frac{\theta(x-x')}{x'' (x-x'')(x''-x')(1-x'')}
\end{eqnarray}
with $s_1,s_2,s_3$ given by:
\begin{eqnarray*}
s_1&=&E_{a'}+E_{c'}+E_{b'''},\\
s_2&=&E_{a'}+E_{c'}+E_d+E_{b},\\
s_3&=&E_{a''}+E_{d}+E_{b}.
\end{eqnarray*}

%%%%%%%%%%%%%%%%%%%%%%%%%%%%%%%%%%%%%%%%%%%%%%%%%%%%%%%

\end{document}